\documentclass[12pt]{article}

\def\beq{\begin{equation}}
\def\eeq{\end{equation}}
\def\beqn{\begin{eqnarray}}
\def\eeqn{\end{eqnarray}}
\def\lsim{\mathrel{\rlap{\lower3pt\hbox{\hskip0pt$\sim$}}
    \raise1pt\hbox{$<$}}}
\def\gsim{\mathrel{\rlap{\lower4pt\hbox{\hskip1pt$\sim$}}
    \raise1pt\hbox{$>$}}}

\begin{document}


\begin{flushright}
ITEP-TH-55/02\\
hepth/0210281\\

\end{flushright}

\vspace{1.3cm}

\begin{center}
\baselineskip25pt

{
\Large\bf

Konishi anomaly and N=1 effective superpotentials from the matrix models
  }

\vspace{0.3cm}

{\bf  A. Gorsky}

 {\em Institute of Theoretical and Experimental Physics,
B.Cheremushkinskaya 25, Moscow,  117259, Russia }

\vspace{0.1cm}

\end{center}

\vspace{1cm}
\begin{center}
{\large\bf Abstract} \vspace*{.25cm}
\end{center}
We discuss the restrictions imposed by the
Konishi anomaly  on the matrix
model approach to the calculation of the effective
superpotentials in N=1 SUSY gauge theories
with different matter content. It is shown that they correspond to
the anomaly deformed Virasoro  $L_0$ constraints .


\vspace{1cm}


1. Recently \cite{dv} it was suggested that
the effective
Veneziano-Yankielowicz type superpotentials for the composite
glueball chiral superfield $S= \frac{1}{32\pi^2}Tr W_{\alpha}W^{\alpha}$
in N=1 SUSY theories can be evaluated using
the planar limit of the matrix models.
Such superpotentials  provide the
tensions of the BPS domain walls which
saturate the central charges in SUSY algebra
and therefore enjoy the known properties
under the RG flows. This approach based on the earlier
analysis in \cite{cv}
has been considered for the N=1 theories with several matter
fields in adjoint \cite{dorey}. Recently it was  argued
\cite{ferrari} that
planar limit in the matrix model corresponds to the
Intriligator-Leigh-Seiberg nonrenormalization conjecture
\cite{ils}
for the generic superpotentials.
The matrix model approach has been  also analyzed in \cite{chekhov,fo,dgkv}.
To
clarify the validity of the approach it is interesting
to derive the full set of the field theoretic constraints
imposed by the symmetries of the problem and check
their consistency with the matrix model realization.

In this note we formulate the constraint on the
superpotential imposed by the Konishi anomaly.
Such Konishi anomaly relation has been proven very useful
in the derivation of the gluino condensate since
it allows to perform the transition from the weak
to strong coupling regime in the controllable way \cite{svz}.
Recently it was also demonstrated that it provides
the effective bridge between N=2 and N=1 SUSY theories
with some matter content \cite{gvy}.

We shall argue that Konishi relation is fulfilled
in the theory with the single field in adjoint
and in the  elliptic models with the proper normalizations
of the condensates.
From the matrix model point of view these relations
can be interpreted as a version of $L_0$ constraint.

2.
We would like to discuss the following
chiral matter superfield transformation
\beq
\Phi_f(x_{L},\theta) \rightarrow e^{-i\alpha_f}\Phi_f(x_{L},\theta)
\label{transform}
\eeq
Contrary to the R-symmetry transformation the
transformation (\ref{transform}) does not touch the $\theta$ variable.
The corresponding Konishi current
\beq
J_f= \bar{\Phi}_f e^V \Phi_f
\eeq
suffers from the anomaly
and  the following equation \cite{konishi} takes place
\beq
\bar{D}^2\bar{\Phi _f}e^{V} \Phi_f = \frac{T(f)}{2\pi ^2} Tr W^2 +
4Tr\Phi_f W^{'}(\Phi_f)
\label{equation}
\eeq
where $W(\Phi)$ is the tree superpotential for the matter field
$\Phi$ and the value of the dual Coxeter number in the matter
representation $T(f)$ enters the anomaly term.
For SU(N) it takes values $T(adj)=N$ and $T(fund)=1/2$.
Note that in contrast to the anomaly in R-current there is
no higher order correction to (\ref{equation}).

The key point
amounting to the importance of the Konishi anomaly relation
follows from the vanishing of the l.h.s upon the averaging
over the supersymmetric vacuum state. Hence one arrives
at the following relation between condensates in N=1 theory
involving fields $\Phi_{i}$ in representations $R_{i}$
\beq
0=4T(R_i) <S> + <Tr\Phi W^{'}(\Phi_i)>
\eeq
Note that we have such equation for each matter field.
Let us emphasize that composite glueball field is uncharged
with respect to the Konishi transformation while it is
natural to assign charge k to the coupling $t_k$ in the
superpotential $W=\sum _{k} t_kTr\Phi^k$.

We shall be interested in the low energy effective superpotentials
hence let us formulate the equation on the $W_{eff}$ following
from the Konishi relation in the theory with the single adjoint field.
To this aim remind that
\beq
\frac{dW_{eff}}{dt_k} =<Tr \Phi^k>
\eeq
and
\beq
1/2\frac{dW_{eff}}{dLog \Lambda^{2N}} =-<S>
\eeq
therefore one arrives at
\beq
(\sum _{k}kt_k \frac{d}{dt_k} -\frac{d}{d log\Lambda})<W_{eff}(t_k,\Lambda)>=0
\label{constraint}
\eeq
where the superpotential is taken at some vacuum state.

Let us identify (\ref{constraint}) in the matrix model approach.
According to \cite{dv} the effective superpotential can be obtained from
the tree one via the following procedure.
For example, the starting point
in the theory with the single adjoint is
the integral over the Hermitian matrixes
\beq
Z=\frac{1}{Vol[\Phi]}\int d\Phi e^{-\frac{1}{g_s}W(\Phi)}
\eeq
with the tree superpotential. At large N limit one can derive
the density of the eigenvalues in the saddle point approximation
and calculate the partition sum in the planar limit.
Given the matrix
model partition function the effective
superpotential can be derived from the following
formula
\beq
W_{eff}(S,t_k)= NSlogS/\Lambda^3 - 2\pi i \tau S +
N \frac{\partial {\cal{F}} _{0}(S,t_k)}{\partial S}
\eeq
where ${\cal{F}}_0$ is the free energy of the matrix model
in the planar limit.
Minimization of $W_{eff}$ with respect to the composite  field
S which is identified with $S= g_s N$ in the
matrix model amounts to the vacuum values of the
superpotential yielding the
holomorphic physical observables - domain
wall tensions.

Note that there are essentially two different contributions
to the effective superpotential. One yielding the entropy
contribution to the free energy
amounts from the volume of the group while the second comes
from the normalized matrix integral. In what follows we
shall argue that the regularized volume has to respect
the condition follows from the Konishi anomaly.

In terms of the matrix model the Konishi relation
can be reformulated as a anomaly modified Virasoro $L_0$ constraint.
namely
\beq
L_0 W_{eff}(\Lambda,t_k) = 4T(R)<S>
\eeq
The Virasoro operator $L_0$ term is wellknown in the matrix models and
reflect the invariance of the matrix integral under the
change of variables. Let us note that it is essential
to consider the complex matrix to identify properly
then multiplication by the complex phase corresponding
to the Konishi transformations. The second term
actually reflects noninvariance of the entropy term in the
effective action. It can be also expressed in terms of the
derivative of $W_{eff}$ with respect to $\log \Lambda$ for
the asymptotically free theory or with respect to the
bare coupling $\tau$ in the perturbed finite theory.

3. Let us discuss  a few explicit examples starting with
the theory with the single matter field in adjoint. Consider
first the simplest superpotential $W=\mu Tr\Phi^2$.
The vacuum value of the superpotential
reads as
\beq
W_{min}= \mu \Lambda^2
\eeq
which evidently obeys equation (\ref{constraint}). The next example
involves the cubic tree superpotential $W = \mu Tr\Phi^2 +gTr\Phi^3$.
The corresponding effective superpotential looks as follows \cite{cv,fo}
\beq
W_{eff}= NSlog \frac{S}{\Lambda_{eff}^3} -2\pi i \tau S +
\frac{N}{2}\sum_{k} (\frac{8g^2}{\mu^3})^k \frac{S^{k+1}
\Gamma(3k/2)}{(k+1)!\Gamma(k/2+1)}
\eeq
Since S is uncharged we have to check the Konishi constraint for
both terms separately. The sum evidently obey the constraint as for the
entropy term we have to assume that
\beq
\Lambda_{eff}^3= \mu \Lambda^2
\eeq
Such identification is natural from the matrix models
indeed.

Let us turn to the softly broken N=4 theory and assume that
masses of adjoint scalars are different. On the field theory
side we immediately derive the system of three equation
on the condensates
\beq
- 4T(adj)<S>= <iTr\Phi_1[\Phi_2,\Phi_3]> +2 M_k<Tr \Phi_k ^2>
\label{adjoint}
\eeq
where $k=1,2,3$. It is evident that $M_k<Tr \Phi_k ^2>$ does not
depend on the flavor. To compare (\ref{adjoint}) with the
matrix model answer recall that in this theory gluino condensate
can be derived from the superpotential as follows
\beq
<S>= -\frac{1}{2\pi i}\frac{dW_{eff}}{d\tau}
\eeq
where $\tau$ is the complexified bare coupling constant.

The effective superpotential on the field theory side
which effectively sums
the contributions from the fractional instantons
has been suggested in \cite{dorey} and in the
k-th confining vacuum looks as follows
\beq
W_{eff}=M_1 M_2 M_3  \frac{N^3}{24}[E_2(\tau) - \frac{1}{N}E_2(\frac{\tau +k}{N})
+A(\tau,N))]
\eeq
where $A(N,\tau)$ is some unknown holomorphic function
of $\tau$ which does not depend on
 k and $E_2(\tau)$ is the regulated second Eisenstein series
\beq
E_2= \frac{3}{\pi^2} \sum_{n,m} \frac{1}{(m+n\tau)^2}.
\eeq

The matrix model answer has been discussed in \cite{dorey}
where it was found that in the vacua corresponding to
$\Phi_i=0$ classical configuration the superpotential
differs from the field theory answer by the additive constant.
To check the Konishi relation we have to calculate the
vacuum expectation value of the operator $<Tr\Phi_1[\Phi_2,\Phi_3]>$
independently. To this aim let us consider
more general model involving additional couplings \cite{dhk}
with the tree superpotential
\beq
W_{\beta,\lambda}=Tr(i\lambda\Phi[\Phi_{+},\Phi_{-}]_{\beta} +M\Phi_{+}\Phi_{-}
+\mu\Phi^2)
\eeq
where the Leigh-Strassler deformation \cite{ls} is considered
\beq
[\Phi_{+},\Phi_{-}]_{\beta}=\Phi_{+}\Phi_{-} e^{i\beta/2} - \Phi_{-}\Phi_{+} e^{-i\beta/2}
\eeq

The matrix partition function for this model
\beq
Z=\int d\Phi_{-} d\Phi_{+} d\Phi e^{-\frac{W_{\beta,\lambda}}{g_s}}
\eeq
can be done in the saddle point approximation
using \cite{kostov} amounting to the effective
superpotentials in the massive $(p,k)$ vacuum states \cite{dhk}
\beq
W_{eff}=\frac{pN\mu M^2}{2\lambda^2 sin\beta}\times
\frac{\theta_{1}^{'}(p\beta/2|\tilde{\tau})}{\theta_{1}(p\beta/2|\tilde{\tau})}
\label{kostov}
\eeq
where
\beq
\tilde{\tau}=\frac{p(\tau +iNlog\lambda/\pi) +k}{q}
\eeq
where $k=0,1, \dots,q-1$ and
the following representation is known
\beq
\frac{\theta_{1}^{'}(x|\tau)}{\theta_{1}(x|\tau)}=
cotx +4\sum _{n=1} \frac{q^{2n}sin2nx}{1-q^{2n}}
\eeq

Now we can calculate relevant condensates just
from derivatives of $W_{eff}$ with respect to
couplings $\mu, M,\beta$ and $\lambda$.  This can be most
easily seen by taking the simplified case
$\beta=0$. In this case we can calculate condensate
\beq
<Tr\Phi_1[\Phi_2,\Phi_3]>=-i\frac{d W_{eff}}{d\lambda}
\eeq
with derivative taken at $\lambda=1$.
Collecting all derivatives of the superpotential in the
vacuum state we see that Konishi relation
\beq
(-\frac{Ni}{\pi}\frac{d}{d\tau} + \frac{d}{d\lambda}_{|\lambda=1} +
2\mu \frac{d}{d\mu} )W_{eff}(\lambda,\mu,\tau)=0
\eeq
is fulfilled. \footnote{I am grateful to N.Dorey, T.Hollowood and
S.P.Kumar who showed that the statement on the contradiction
with the Konishi relation for eliptic model claimed in the earlier
version is based on the wrong normalization
of the condensate}

One more example involves
softly broken N=2 theory with  the matter fields in the fundamental.
The field theory predicts system of $N_f+1$ equations for condensates
\beq
4T(f)<S> =< Tr \tilde{Q_i} \Phi Q_i> + m_i<Tr \tilde{Q_i}Q_i>
\eeq
\beq
4T(a)<S> = \sum_ i <Tr \tilde{Q_i} \Phi Q_i> +
\sum_k kt_k<Tr\Phi^k>
\eeq
where $i=1,...,N_f$. The linear combination of the equations does not
contain the Yukawa terms
\beq
4(T(a)-N_fT(f))<S>= -\sum_{i}m_i<Tr \tilde{Q_i}Q_i> +\sum_k kt_k<Tr\Phi^k>
\label{fund}
\eeq
and can be translated into condition for the superpotential
\beq
(- \sum _{i} m_i\frac{d}{dm_i} + \sum _{k}kt_k \frac{d}{dt_k} -
\frac{d}{d log\Lambda})<W_{eff}(t_k,\Lambda,m_i)>=0
\eeq
Note that in the perturbed superconformal theory with $N_f= 2N_c$
anomaly contribution in (\ref{fund}) vanishes.

Note that in the theory with fundamentals there are also additional
constraints amounting from the Ward identities \cite{veneziano}.
In particular in the theory with large adjoint mass when it
effectively decouples the following relation takes place
\beq
<S>=Nm\frac{d}{dm}<S>
\eeq
which claims that $<S>$ is the holomorphic function
of $m$ and $<S>\propto m^{1/N}$.

4.
In this note we discussed the constraint imposed by the Konishi anomaly
on  the vacuum values of the superpotentials calculated within the
matrix model approach. We focused on the single cut solution in
the matrix models corresponding to the simplest gauge group splitting.
It was argued that the constraint can be considered
in the matrix model as some
version of the Virasoro constraint modified by anomaly term.
We have tested several models with respect to constraint. It appeared
that it is satisfied in the N=1 theory with the single adjoint field
and in the elliptic models.
The condition for the superpotential in the theory with the fundamental matter
is presented.

A few additional comments are in order. First let us note that one
could expect additional constraints imposed on the domain wall tensions or
equivalently on the vacuum values of superpotentials. Actually
these should be formulated as a kind of Picard-Fuchs equations
for the integrals of the  forms over the corresponding
cycles. Indeed it is known how the domain wall tensions can
be calculated in terms of such integrals
of the holomorphic three forms \cite{cv}. It is natural
to expect the whole multiplet of the domain wall tensions
considered as a function on couplings obey some
single higher order differential Picard-Fuchs equation
which has singularities at Argyres-Douglas points where
collisions of vacua happen and vanishing cycles emerge.
On the other hand it would be interesting to realize the hypothetical
symmetry meaning of the higher Virasoro constraints evident in the
matrix model on the field theory side.

There is also some analogy with the low energy description
of the nonsupersymmetric QCD in terms of chiral lagrangian.
The matrix integral over the flavor unitary matrixes in QCD
contains the important information concerning the order
parameters of the low energy theory. The reason for the
matrix model to work is that  it captures
the information about the spectrum of the Dirac operator
in the complicated instanton ensemble background
and such matrix models mimics the integration over the
instanton moduli space. For instance, the counterpart
of the expression for the gluino condensate in terms of the
integrals of the spectral density in the matrix model
is the Casher-Banks relation for the chiral condensate in QCD
\beq
<\bar{\Psi}\Psi>=const\rho(0)
\eeq
where $\rho(0)$ is the value of the spectral density of the Dirac
operator at origin. Since the matrix model relevant for N=1 SYM
has the interpretation in terms of the ADHM construction for
D-instantons \cite{kkn} it would be interesting to pursue
this analogy further.

One more comment concerns the interpretation in terms
of the classical integrable many-body systems.
The vacuum states in N=1 SYM theory corresponds to the
equilibrium states in the corresponding integrable many body system
\cite{dorey1}(see \cite{gm} for review).
The point we would
like to mention is that  fermions arising in the matrix model
are related to the eigenfunctions of the Lax operator in the
integrable system. The physical model which provides
the additional intuition concerning meaning of the corresponding
spectral curve is the Peierls model of one-dimensional
superconductivity. In particular one can map the generation
of the scale in the N=1 SYM theory into the gap formation
for fermions in Peierls type models \cite{gor}.

Finally it is interesting to question if the matrix model
could provide information concerning another BPS objects
existing in N=1 theories. Actually there are two types
of BPS objects in N=1 SYM theory saturating central
charges in N=1 SUSY algebra. Apart from the domain
walls saturating the central charge in $\{Q,Q\}$
there are strings or domain wall junctions
saturating central terms in $\{\bar{Q},Q\}$. To
discuss junctions one has to care on both central
charges \cite{gs} and it would be interesting to understand
whether matrix model  could capture the information
about the objects with 1/4 instead of 1/2 amount
of SUSY.

I am grateful to A. Marshakov, A. Mironov and A. Morozov for the
useful comments.
I would like to thank N.Dorey, T.Hollowood
and S.P.Kumar for  pointing out
the wrong normalization of the
condensate in the elliptic model in the earlier version of the paper.
The work was supported in part by grant
INTAS-00-334.


\begin{thebibliography}{99}

\bibitem{dv}
R.~Dijkgraaf and C.~Vafa,
arXiv:hep-th/0208048 \\
R.~Dijkgraaf and C.~Vafa,
Nucl.\ Phys.\ B {\bf 644}, 21 (2002)
[arXiv:hep-th/0207106]. \\
R.~Dijkgraaf and C.~Vafa,
Nucl.\ Phys.\ B {\bf 644}, 3 (2002)
[arXiv:hep-th/0206255].

\bibitem{cv}
F.~Cachazo, K.~A.~Intriligator and C.~Vafa,
Nucl.\ Phys.\ B {\bf 603}, 3 (2001)
[arXiv:hep-th/0103067]. \\
F.~Cachazo and C.~Vafa,
arXiv:hep-th/0206017.


\bibitem{dorey}
N.~Dorey, T.~J.~Hollowood, S.~Prem Kumar and A.~Sinkovics,
arXiv:hep-th/0209089. \\
N.~Dorey, T.~J.~Hollowood, S.~P.~Kumar and A.~Sinkovics,
arXiv:hep-th/0209099.

\bibitem{ferrari}
F.~Ferrari,
arXiv:hep-th/0210135.
\bibitem{ils}
K.~A.~Intriligator, R.~G.~Leigh and N.~Seiberg,
Phys.\ Rev.\ D {\bf 50}, 1092 (1994)
[arXiv:hep-th/9403198].
\bibitem{chekhov}
L.~Chekhov and A.~Mironov,
arXiv:hep-th/0209085.
\bibitem{fo}
H.~Fuji and Y.~Ookouchi,
arXiv:hep-th/0210148.
\bibitem{dgkv}
R. Dijkgraaf, S. Gukov, V.Kazakov and C. Vafa,
arXiv:hep-th/0210238
\bibitem{svz}
~A.~Shifman and A.~I.~Vainshtein,
Nucl.\ Phys.\ B {\bf 227}, 456 (1986) \\
~A.~Shifman and A.~I.~Vainshtein,
Nucl.\ Phys.\ B {\bf 296}, 445 (1988)
\bibitem{gvy}
A.~Gorsky, A.~I.~Vainshtein and A.~Yung,
Nucl.\ Phys.\ B {\bf 584}, 197 (2000)
[arXiv:hep-th/0004087].

\bibitem{konishi}
K.~Konishi,
Phys.\ Lett.\ B {\bf 135}, 439 (1984).
\bibitem{dhk}
N. Dorey, T. Hollowood and S. Kumar,
arXiv:hep-th/0210239
\bibitem{ls}
R.~G.~Leigh and M.~J.~Strassler,
Nucl.\ Phys.\ B {\bf 447}, 95 (1995)
[arXiv:hep-th/9503121].
\bibitem{kostov}
I.~K.~Kostov,
Nucl.\ Phys.\ B {\bf 575}, 513 (2000)
[arXiv:hep-th/9911023].

\bibitem{veneziano}
D. Amati, G. Rossi and G. Veneziano,
Nucl.\ Phys.\ B {\bf 249}, 1 (1985)

\bibitem{kkn}
V.~A.~Kazakov, I.~K.~Kostov and N.~A.~Nekrasov,
Nucl.\ Phys.\ B {\bf 557}, 413 (1999)
[arXiv:hep-th/9810035].
\bibitem{dorey1}
N.~Dorey,
JHEP {\bf 9907}, 021 (1999)
[arXiv:hep-th/9906011].
\bibitem{gm}
A.~Gorsky and A.~Mironov,
arXiv:hep-th/0011197.
\bibitem{gor}
A.~Gorsky,
Mod.\ Phys.\ Lett.\ A {\bf 12}, 719 (1997)
[arXiv:hep-th/9605135].

\bibitem{gs}
A.~Gorsky and M.~A.~Shifman,
Phys.\ Rev.\ D {\bf 61}, 085001 (2000)
[arXiv:hep-th/9909015].
\end{thebibliography}
\end{document}